\documentclass[12pt]{article}
\usepackage{epsfig}
\usepackage{amsfonts}
\usepackage{latexsym}
\usepackage{amsmath}
\usepackage{mathrsfs}
\usepackage{hyperref}
\usepackage{setspace}
\usepackage{color}
\usepackage{bm}
\usepackage{slashed}
\numberwithin{equation}{section}

\begin{document}

\begin{titlepage}
\unitlength = 1mm
%\today 
\begin{flushright}
KOBE-COSMO-17-08
\end{flushright}

\vskip 1cm
\begin{center}

{\Large {\textsc{\textbf{Infinite violation of Bell inequalities in inflation}}}}

\vspace{1.8cm}
Sugumi Kanno$^{*\,\flat}$ and Jiro Soda$^{\natural}$ 

\vspace{1cm}

\shortstack[l]
{\it $^*$ Department of Theoretical Physics and History of Science,
University of the Basque Country\\
~~48080 Bilbao, Spain\\
\it $^\flat$ IKERBASQUE, Basque Foundation for Science, 
Maria Diaz de Haro 3,
48013, Bilbao, Spain\\
\it $^\natural$ Department of Physics, Kobe University, Kobe 657-8501, Japan
}

\vskip 1.5cm

\begin{abstract}
\baselineskip=6mm
We study the violation of Bell-Mermin-Klyshko (BMK) inequalities in initial quantum states of scalar fields in inflation. We show that the Bell inequality is maximally violated by the Bunch-Davies vacuum which is a two-mode squeezed state of a scalar field. However, we find that the violation of the BMK inequalities does not increase with the number of modes to measure. We then consider a non-Bunch-Davies vacuum expressed by a four-mode squeezed state of two scalar fields. Remarkably, we find that the violation of the BMK inequalities increases exponentially with the number of modes to measure. This indicates that some evidence that our universe has a quantum mechanical origin may survive in CMB data even if quantum entanglement decays exponentially afterward due to decoherence.

\end{abstract}

\vspace{1.0cm}

\end{center}
\end{titlepage}

\pagestyle{plain}
\setcounter{page}{1}
\newcounter{bean}
\baselineskip18pt

\setcounter{tocdepth}{2}

\tableofcontents

\section{Introduction}
\label{s1}

Einstein's famous phrase as a critique of quantum mechanics ``spooky action at a distance" is recently referred to as ``quantum non-locality". It implies that one particle of entangled pair instantaneously knows what measurement has been performed on the other irrespective of their separation even beyond the lightcone~\cite{Einstein:1935rr}. Several scientists studied local classical hidden variable theories in an attempt to explain the probabilistic nature of quantum mechanics by underlying inaccessible variables. Then Bell derived an inequality that provides a testable difference between the predictions of quantum non-locality and local classical hidden variable theories~\cite{Bell:1964kc,Clauser:1969ny}.
Through sophisticated Bell test experiments, local classical hidden variable theories have been almost ruled out~\cite{Aspect:1981zz, Loop}. 

The Bell inequality is originally formulated for a pair of spins, that is, for a two-partite system. This inequality is violated in the presence of quantum non-locality. Then, how much can the Bell inequality be violated? To answer this question, Tsirelson derived an upper bound on the quantum non-locality later~\cite{Cirelson:1980ry}. The inequality is extended to a multipartite system which is referred to as Bell-Mermin-Klyshko (BMK) inequalities~\cite{Mermin, B_K, Gisin:1998ze}. The quantum upper bound was also generalized for the multipartite system~\cite{Werner, Alsina}. In order to gain some insight into quantum field theories, the BMK inequalities are generalized with continuous quantum variables~\cite{Chen2002, Martin:2016tbd}.
In recent years, more interest has been paid to classifying the multipartite system and quantifying how much we can make use of quantum non-locality in quantum information by using the BMK inequalities~\cite{Nagata,Yu}.

The quantum non-locality should play an important role in cosmology. 
One of the cornerstones of inflationary cosmology is that primordial density fluctuations have a quantum mechanical origin. Hence, the initial state of the universe produced by inflation is highly entangled. It is desired to find compelling evidence for their quantum nature. 
Several studies have been made on quantifying the initial state entanglement by using some measure of entanglement such as the Bell inequality~\cite{Campo:2005sv}, entanglement entropy~\cite{Maldacena:2012xp, Kanno:2014lma, Iizuka:2014rua, Kanno:2016qcc}, entanglement negativity~\cite{Kanno:2014bma} and quantum discord~\cite{Lim:2014uea, Martin:2015qta, Kanno:2016gas}. Recently, Maldacena considered an inflationary scenario where one can prove the quantum origin of density fluctuations by performing the Bell inequality violating experiment during inflation~\cite{Maldacena:2015bha, Choudhury:2016cso}.

In inflationary cosmology, the Bunch-Davies vacuum is usually assumed  as the simplest initial state of quantum fluctuations of the universe. This is because spacetime  looks flat at short distances and then quantum fluctuations are expected to start in a minimum energy state. However, the latest Planck data show the possibility of deviation from the Bunch-Davies vacuum~\cite{Ade:2015lrj}. Motivated by this, there have been several attempts to find some observational signatures on the CMB when the initial state is a non-Bunch-Davies vacuum due to entanglement between two scalar fields~\cite{Albrecht:2014aga, Kanno:2015ewa}, between two universes~\cite{Kanno:2015lja}, and due to scalar-tensor entanglement~\cite{Collins:2016ahj,Bolis:2016vas}. If we apply the BMK inequalities violating experiment to cosmology, we may be able to prove the quantum origin of density fluctuations and find the nature of the initial state of the universe. 

In this paper, we evaluate the BMK inequalities for the Bunch-Davies vacuum and a non-Bunch-Davies vacuum in inflation. We find that both vacua violate the BMK inequalities.  Remarkably, as for the non-Bunch-Davies vacuum, the violation increases exponentially with the number of modes to measure. This indicates the detection of the quantum non-locality of primordial density fluctuations. 

The paper is organized as follows. In section 2, we review Bell and Mermin-Klyshko inequalities and explain how to classify the quantum non-locality by using the BMK inequalities. We then introduce pseudo-spin operators in order to extend 
the BMK inequalities to quantum field theories.
In section 3, as cosmological initial states, we explain the Bunch-Davies vacuum expressed by a two-mode squeezed state and the non-Bunch-Davies vacuum expressed by a four-mode squeezed state. In section 4, we evaluates the BMK inequalities for those cosmological initial states. Finally we summarize our result and discuss the implications in section 5.

\section{Bell-Mermin-Klyshko inequalities}
\label{s2}

In this section, we review Bell inequality with the simplest example of a pair of spins (a two-partite system) and Mermin-Klyshko (BMK) inequalities for a multipartite system ~\cite{Bell:1964kc,Clauser:1969ny}. The BMK inequalities are violated by quantum non-locality and provide a criterion for descriminating the quantum non-locality from any local classical hidden variable theories~\cite{Mermin, B_K}. The upper bound of the violation increases with the number of partite states~\cite{Werner, Alsina} and the entangled states are classified by using the upper bound~\cite{Yu}. 
The pseudo-spin operators are introduced for continuous quantum variables~\cite{Chen2002}. 

\subsection{Bell inequality}
\label{s2-1}

We consider two sets of non-commuting operators $A$, $A'$ and $B$, $B'$. Those operators correspond to measuring the spin along various axes and have eigenvalues $\pm 1$. They are expressed by the Pauli matrices $\sigma^i$ and unit vectors $n^i$ such as $A=n^i\sigma^i$. 
The Bell operator ${\cal B}$ is defined as
\begin{eqnarray}
{\cal B}=\frac{1}{2} \left( A\otimes B+A'\otimes B+A\otimes B'-A'\otimes B'  \right)\,,
\label{bell1}
\end{eqnarray}
where the variables $A$, $A'$ and $B$, $B'$ are represented by Hermitian operators which act on the Hilbert spaces ${\cal H}_A$ and ${\cal H}_B$ respectively. If we rewrite it as a factorized form
\begin{eqnarray}
{\cal B}=\frac{1}{2}A\otimes \left(B+B'\right)+\frac{1}{2}A'\otimes\left(B-B'\right)\,,
\label{bell2}
\end{eqnarray}
then we see that the first (second) term becomes $\pm 1$ while the second (first) one vanishes because we can have either $B=B'$ or $B=-B'$. In local classical hidden variable theories, the expectation value of ${\cal B}$ then gives $|\langle{\cal B}\rangle|\leq 1$. In quantum mechanics, however, this Bell inequality  can be violated for the expectation value of the quantum operator.
It is easy to check that its square becomes\footnote{The tensor product $\otimes$ is omitted below for simplicity unless there may be any confusion.}
\begin{eqnarray}
{\cal B}^2=I- \frac{1}{4}\left[A,A'\right]\left[B,B'\right]\,,
\end{eqnarray}
where we used the fact that the square of each operator is one, $A^2=I$, $A'^2=I$, etc and $I$ is the identity operator. Since the commutators of the Pauli matrices are non-zero\footnote{The Pauli matrices satisfy $[\sigma_a,\sigma_b]=2i\varepsilon_{abc}\,\sigma_c$ where $\varepsilon_{abc}$ is antisymmetric tensors. For local classical hidden variable theories, the commutators are zero and $\langle{\cal B}^2\rangle\leq 1$ or $|\langle{\cal B}\rangle|\leq 1$.} and each gives $2i$, we find that $\langle{\cal B}^2\rangle\leq 2$ or $|\langle{\cal B}\rangle|\leq \sqrt{2}$. Thus the maximal violation of Bell inequality in quantum mechanics has the extra $\sqrt{2}$ factor in the case of a pair of spins~\cite{Cirelson:1980ry}.

\subsection{Mermin-Klyshko inequalities}
\label{s2-2}

The Bell inequality is generalized for a multipartite system, which is called Mermin-Klyshko inequalities. We write the operators $\{A,B,C,\cdots\}$ by $\{{\cal O}_1,{\cal O}_2,{\cal O}_3,\cdots\}$ below for later convenience. Defining ${\cal B}_1={\cal O}_1$ and ${\cal B}'_1={\cal O}'_1$, the Mermin-Klyshko operator is defined recursively as
\begin{eqnarray}
{\cal B}_n=\frac{1}{2}\,{\cal B}_{n-1}\left({\cal O}_n+{\cal O}'_n\right)
+\frac{1}{2}\,{\cal B}'_{n-1}\left({\cal O}_n-{\cal O}'_n\right)\,,\qquad n=2,3,4,\cdots
\label{MK1}
\end{eqnarray}
where ${\cal B}'_{n-1}$ is obtained from ${\cal B}_{n-1}$ by interchanging primed and nonprimed operators ${\cal O}_n$. Thus, given the initial terms ${\cal B}_1={\cal O}_1$ and ${\cal B}'_1={\cal O}'_1$, each subsequent term is determined by this relation. Explicitly, the recurrence yields operators 
\begin{eqnarray}
{\cal B}_{2}&=&\frac{1}{2}\,{\cal O}_{1}\left({\cal O}_{2} + {\cal O}'_{2}\right)  
+\frac{1}{2}\,{\cal O}'_{1}\left({\cal O}_{2} - {\cal O'}_{2}\right)\,,\nonumber\\
{\cal B}_3&=&\frac{1}{2}\,{\cal B}_{2}\left({\cal O}_3+{\cal O}'_3\right)
+\frac{1}{2}\,{\cal B}'_{2}\left({\cal O}_3-{\cal O}'_3\right)\,,
\label{B2B3}
\end{eqnarray}
and so on. In local classical hidden variable theories, the Mermin-Klyshko inequalities reads
\begin{eqnarray}
|\langle{\cal B}_n\rangle| \leq 1\,, \qquad n=1,2,3,\cdots\,,
\end{eqnarray}
because we can have ${\cal O}_n={\cal O}'_n$ or ${\cal O}_n=-{\cal O}'_n$.
In quantum mechanics, this inequality is violated and the expectation value of ${\cal B}_n$ can be bigger. In fact, the Mermin-Klyshko inequalities tells~\cite{Werner, Alsina}
\begin{eqnarray}
|\langle{\cal B}_n\rangle| \leq 2^{\frac{n-1}{2}} \,, \qquad n=1,2,3,\cdots\,.
\label{maximal1}
\end{eqnarray}
Thus, in quantum mechanics, the upper bound can be exponentially bigger for multipartite states $(n>2)$.

For later purpose, it is useful to note that the Mermin-Klyshko operators have the following relation~\cite{Gisin:1998ze}
\begin{eqnarray}
{\cal B}_n=\frac{1}{2}\,{\cal B}_{n-p}\left({\cal B}_p+{\cal B}'_p\right)
+\frac{1}{2}\,{\cal B}'_{n-p}\left({\cal B}_p-{\cal B}'_p\right) \,,\qquad n=2,3,4,\cdots\,,
\label{MK2}
\end{eqnarray}
where $p$ is an integer in the range $1\leq p \leq n-1$.
This can be proved by induction from the definition (\ref{MK1}).
For example, we use the following relations later in section~\ref{s4-2}.
\begin{eqnarray}
{\cal B}_8&=&\frac{1}{2}\,{\cal B}_{4}\left({\cal B}_4+{\cal B}'_4\right)
+\frac{1}{2}\,{\cal B}'_{4}\left({\cal B}_4-{\cal B}'_4\right)\,,  \nonumber\\
{\cal B}_{12}&=&\frac{1}{2}\,{\cal B}_{8}\left({\cal B}_4+{\cal B}'_4\right)
+\frac{1}{2}\,{\cal B}'_{8}\left({\cal B}_4-{\cal B}'_4\right)\,.
\label{812}
\end{eqnarray}

\subsection{Classification of entanglement with BMK inequalities}
\label{s2-3}

What kind of entangled state makes the violation of BMK inequalities bigger?
The upper bound of the BMK inequalities is classified as follows~\cite{Nagata,Yu}
. If we introduce the quadratic form of Mermin-Klyshko inequalities, the violation of BMK inequalities Eq.~(\ref{maximal1}) is written as
\begin{eqnarray}
\langle{\cal B}_N\rangle^2+\langle{\cal B}'_N\rangle^2\leq 2^E\,,
\label{quadratic}
\end{eqnarray}
where $N$ is the total number of partite states (Hilbert spaces) and
\begin{eqnarray}
E=N-K_1-2L+1\,,
\label{E}
\end{eqnarray}
where $K_1$ is the number of single separated partite state which is not entangled with other $N-1$ partite states. Let $K_p$ be the number of groups consists of $p$ entangled partite states. Then $L$ is the sum of $K_p$ defined by $L=\sum_{p=2}^{M}K_p$
where different groups are not entangled each other,
and $M$ is the largest number of entangled partite states in a group.
The total number of partite states $N$ is then divided into $N=\sum_{p=1}^{M}pK_p$.

If we use the quadratic form of Bell inequality, the violation of Bell inequality of a pair of spins, $|\langle{\cal B}_2\rangle|\leq \sqrt{2}$, is expressed as $\langle{\cal B}_2\rangle^2+\langle{\cal B}'_2\rangle^2\leq 2$ where $N=2$, $L=1$, $K_1=0$.
As we mentioned in Eq.~(\ref{maximal1}), it is expected that the upper bound can be exponentially increased for multipartite states $n>2$. Let us increase the number of pairs up to $m$ pairs of partite states, $N=2m$, $L=m$, $K_1=0$. Then we see $\langle{\cal B}_{2m}\rangle^2+\langle{\cal B}'_{2m}\rangle^2\leq 2$ still holds and the upper bound of the violation does not increase. This is because both cases hold $N-2L=0$. If we try to consider the case $N-2L\neq 0$, we may get large violation of BMK inequalities.

\subsection{Pseudospin operators}
\label{s2-4}

In order to discuss BMK inequalities in the context of cosmology later, we need to express it in terms of continuous quantum variables. In this section, we introduce pseudospin operators that behave in the same manner as the usual spin $1/2$ operators but the pseudospin operators can be used for continuous quantum variables~\cite{Chen2002}. The pseudospin operators distinguish between even parity and odd parity. 

The pseudospin operators are defined as follows.
The eigenvectors of the pseudospin operator $S_z$ are $|2n+1\rangle$ and $|2n\rangle$. The corresponding eigenvalues are $+1$ and $-1$. The states $|2n+1\rangle$ and 
$|2n\rangle$ are orthogonal to each other, $\langle 2n+1|\,2n\rangle=0$. Thus we define
\begin{eqnarray}
S_z&=&\sum_{n=0}^\infty\Bigl(\,|2n+1\rangle\langle 2n+1|-|2n\rangle\langle 2n|\,\Bigr)\,,\\
S_+&=&\sum_{n=0}^\infty|2n+1\rangle\langle 2n|=S_-^\dag\,,
\end{eqnarray}
where $S_+$ and $S_-$ are the parity-flip operators. The other two components of pseudospin satisfy $S_x\pm iS_y=2S_\pm$. The commutation relations are $[S_+,S_-]=S_z$ and $[S_z,S_\pm]=\pm 2S_\pm$. $S_z^2=S_x^2=S_y^2=I$.
 Then we have
\begin{eqnarray}
&&S_z|2n\rangle=-|2n\rangle\,,\qquad\quad S_z|2n+1\rangle=|2n+1\rangle\,,\nonumber\\
&&S_+|2n\rangle=|2n+1\rangle\,,\qquad S_+|2n+1\rangle=0\,,\nonumber\\
&&S_-|2n\rangle=0\,,\qquad\qquad\quad S_+|2n+1\rangle=|2n\rangle\,.
\label{rules}
\end{eqnarray}
In this way, pseudospin operators act on the Hilbert spaces.
As we will see in section~\ref{s4-1}, the BMK inequalities are generalized with continuous quantum variables of pseudospin operators.

\section{Cosmological initial states and particle creation}
\label{s3}

In order to run the Bell experiment, one repeats the experiment many times on the same quantum state. Thus, to perform a Bell type experiment in cosmology, we need to choose a quantum state in the universe. In this section, we review the Bunch-Davies vacuum expressed by a two-mode squeezed state and a non-Bunch-Davies vacuum expressed by a four-mode squeezed state as cosmological initial states.

\subsection{Two-mode squeezed state}
\label{s3-1}　

In quantum field theory, vacuum is not empty and in fact full of virtual particles, which are created and annihilated continuously in entangled pairs. As the universe expands, those virtual particles are released as ordinary particles. This process is calculated by the Bogoliubov transformation between different vacua. 
To see how particle creation can occur in this process, we consider a simple example with a free massless scalar field in an expanding universe. The metric is
\begin{eqnarray}
ds^2=a^2(\eta)\left[-d\eta^2+\delta_{ij}dx^idx^j\,\right]\,,
\end{eqnarray}
where $\eta$ is the conformal time, $x^i$ are spatial coordinates, $a(\eta)$ is the scale factor and $\delta_{ij}$ is the Kronecker delta. The indices $(i,j)$ run from $1$ to $3$. If we decompose the scalar field $\phi(\eta, x^i)$ in terms of the Fourier modes as $\phi(\eta, x^i)=\sum_{\bm k}\phi_{\bm k}(\eta)\,e^{i\bm k\cdot\bm x}$, the scalar field is expanded as
\begin{eqnarray}
\phi_{\bm k}(\eta)=a_{\bm k}u_k(\eta)+a_{-\bm k}^\dag u_k^*(\eta)\,,\qquad
\left[a_{\bm k},a_{\bm p}^\dag\right]=\delta_{\bm k,\bm p}\,,
\label{exp1}
\end{eqnarray}
where $k$ is the magnitude of the wave number ${\bm k}$ and $*$ denotes complex conjugation. The mode function $u_k$ satisfies
\begin{eqnarray}
u_k^{\prime\prime}+\left(k^2-\frac{a''}{a}\right)u_k=0\,,
\label{mf}
\end{eqnarray}
where a prime denotes the derivative with respect to the conformal time. As the universe expands, it goes through a transition from de Sitter space to a radiation-dominated era. Suppose that the transition occurs at $\eta=\eta_r>0$, then the scale factor changes as
\begin{eqnarray}
a(\eta)=\left\{
\begin{array}{l}
-\frac{1}{H\left(\eta-2\eta_r\right)}\,,\qquad {\rm for}\quad -\infty<\eta<\eta_r\,,\\
\frac{\eta}{H\eta_r^2}\,,\hspace{1.9cm} {\rm for}\hspace{1cm} \eta_r<\eta\,.
\end{array}
\right.
\end{eqnarray}
Note that $a^{\prime\prime}=0$ for the radiation-dominated era. Eq.~(\ref{mf}) gives the normalized modes which behave like the positive frequency modes in the remote past $u_k^{\rm in}$ and in the radiation-dominated era $u_k^{\rm out}$ respectively of the form
\begin{eqnarray}
\left\{
\begin{array}{l}
u_{k}^{\rm in}(\eta) \equiv \frac{1}{\sqrt{2k}}\left(1-\frac{i}{k\left(\eta-2\eta_r\right)}\right)e^{-ik\left(\eta-2\eta_r\right)}
\,,\qquad {\rm for}\quad -\infty<\eta<\eta_r\,,\\
 u_{k}^{\rm out}(\eta)  \equiv \frac{1}{\sqrt{2k}}\,e^{-ik\eta}\,,\hspace{4.2cm} {\rm for}\hspace{1cm} \eta_r<\eta\,.
\end{array}
\right.
\end{eqnarray}
Then the scalar field Eq.~(\ref{exp1}) is expanded as the following two ways
\begin{eqnarray}
a(\eta)\,\phi_{\bm k}=\left\{
\begin{array}{l}
\int\frac{d^3\bm k}{\sqrt{(2\pi)^3}}\left[
a_{\bm k}^{\rm in}\,u_{k}^{\rm in}+a_{-\bm k}^{\rm in\,\dag}\,u_{k}^{*\,\rm in}
\right]e^{i{\bm k}\cdot{\bm x}}\,,\\
\int\frac{d^3\bm k}{\sqrt{(2\pi)^3}}\left[
a_{\bm k}^{\rm out}\,u_{k}^{\rm out}+a_{-\bm k}^{\rm out\,\dag}\,u_{k}^{*\,\rm out}
\right]e^{i{\bm k}\cdot{\bm x}}\,.
\end{array}
\right.
\end{eqnarray}

Since the positive frequency modes $u_k^{\rm in}$ and $u_k^{\rm out}$ are different, the creation and annihilation operators are different. 
Then the Bunch-Davies vacuum (in-vacuum) $|0_{\rm in}\rangle$ and a vacuum (out-vacuum) $|0_{\rm out}\rangle$ are defined as
\begin{eqnarray}
a_{\bm k}^{\rm in}|0_{\rm in}\rangle=0\,,\qquad
a_{\bm k}^{\rm out}|0_{\rm out}\rangle=0\,.
\label{vacua}
\end{eqnarray}
The initial Bunch-Davies vacuum looks different from the point of view of the out-vacuum.
The relation between these different vacua is expressed by a Bogoliubov transformation:
\begin{eqnarray}
u_{\bm k}^{\rm in}=\alpha_k \,u_{\bm k}^{ \rm out} + \beta_k^* u_{-\bm k}^{\rm out\,\dag}\,,
\end{eqnarray}
or equivalently
\begin{eqnarray}
a_{\bm k}^{\rm in}=\alpha_k^* \,a_{\bm k}^{ \rm out} - \beta_k a_{-\bm k}^{\rm out\,\dag}\,,
\end{eqnarray}
where $\alpha_k$ and $\beta_k$ are Bogoliubov coefficients with $|\alpha_k|^2-|\beta_k|^2=1$. The Bogoliubov coefficients are calculated as
\begin{eqnarray}
\alpha_k&=&\left(u_k^{\rm out},u_k^{\rm in}\right)\Big|_{\eta=\eta_r}
=-\frac{1}{2k^2\eta^2_r}e^{2ik\eta_r}\left(1-2k^2\eta_r^2-2ik\eta_r\right)\,,\\
\beta_k^*&=&-\left(u_k^{* \rm\,out},u_k^{\rm in}\right)\Big|_{\eta=\eta_r}
=-\frac{1}{2k^2\eta_r^2}\,,
\end{eqnarray}
where the Klein-Gordon inner product is defined by $\left(f,g\right)=i\,\bigl\{f^*g^\prime-gf^{*\prime}\bigr\}$. An observer in the out-vacuum will observe particles defined by the operators $a_{\bm k}^{\rm out}$. The expected number of such particles is given by
\begin{eqnarray}
\langle 0_{\rm in}|a_{\bm k}^{\rm out\,\dag}a_{\bm k}^{\rm out}|0_{\rm in}\rangle
=|\beta_k|^2\,.
\end{eqnarray}
This is the creation of particles as a consequence of the cosmic expansion.

Plugging the $a_{\bm k}^{\rm in}$ into the definition of $|0_{\rm in}\rangle$ in Eq.~(\ref{vacua}) and by using 
$[a_{\bm k}^{\rm out}, a_{\bm p}^{\rm out\,\dag}]=\delta_{\bm k,\bm p}$, then the Bunch-Davies vacuum $|0_{\rm in}\rangle$ can be written in terms of $a_{\bm k}^{\rm out\,\dag}$, $a_{-\bm k}^{\rm out\,\dag}$ and the vacua associated to each mode, $|0_{\bm k}^{\rm out}\rangle$ and $|0_{-\bm k}^{\rm out}\rangle$
\begin{eqnarray}
|0_{\rm in}\rangle={\bar N}\exp\left[\sum_{\bm k}\frac{\beta_k}{\alpha_k^*}
a_{\bm k}^{\rm out\,\dag}a_{-\bm k}^{\rm out\,\dag}\right]
|0_{\rm out}\rangle\,,
\label{bogoliubov1}
\end{eqnarray}
where ${\bar N}$ is the normalization factor, and
$|0_{\rm out}\rangle=|0_{\bm k}^{\rm out}\rangle\otimes|0_{-\bm k}^{\rm out}\rangle$. 
This describes a two-mode squeezed state of $n$ pairs of particles since the exponent in Eq.~(\ref{bogoliubov1}) is expanded as
\begin{eqnarray}
|0_{\rm in}\rangle=\prod_{\bm k}\sum_{n=0}^\infty\frac{\tanh^nr_k}{\cosh r_k}\,
|n_{\bm k}^{\rm out}\rangle\otimes|n_{-\bm k}^{\rm out}\rangle\,,
\label{two-mode}
\end{eqnarray}
where ${\bar N}=\prod_{\bm k}\cosh^{-1} r_k$ and a new parameter $r_k$ known as the squeezing parameter is defined as
\begin{eqnarray}
\tanh r_k=\biggl|\frac{\beta_k}{\alpha_k^*}\biggr|=\biggl|\frac{1}{1-2k^2\eta_r^2-2ik\eta_r}\biggr|\,.
\end{eqnarray}
Note that  $r_k \gg 1$  corresponds to the end of inflation ($k\eta_r\ll1$).
We see that the Bunch-Davies vacuum is expressed by a two-mode squeezed state of the modes ${\bm k}$ and $-{\bm k}$.

\subsection{Four-mode squeezed state}
\label{s3-2}

The Bunch-Davies vacuum is usually assumed  as the simplest initial state of quantum fluctuations of the universe. This is because spacetime 
 looks flat at short distances and then quantum fluctuations are expected to start in a minimum energy state. However, the latest Planck data show the possibility of deviation from the Bunch-Davies vacuum~\cite{Ade:2015lrj}. Here, we discuss a four-mode squeezed state as a simple example of non-Bunch-Davies vacua. This state is discussed in~\cite{Albrecht:2014aga, Kanno:2015ewa} with two scalar fields, and also discussed in the context of the multiverse~\cite{Kanno:2015lja}.

We consider two free massive scalar fields $\phi(x^\mu)$ and $\chi(x^\mu)$ in de Sitter space. In Fourier space, they are expanded as
\begin{eqnarray}
\phi_{\bm k}&=&a_{\bm k}^{\rm in}\,u_k^{\rm in}(\eta)
+a_{-\bm k}^{\rm in\,\dag}\,u_k^{*\,\rm in } (\eta)\,,\\
\chi_{\bm k}&=&b_{\bm k}^{\rm in}\,v_k^{\rm in}(\eta)
+b_{-\bm k}^{\rm in \,\dag}\,v_k^{*\,\rm in}(\eta)\,.
\end{eqnarray}
The Bunch-Davies vacuum state is annihilated by both $a_{\bm k}$ and $b_{\bm k}$
\begin{eqnarray}
a_{\bm k}^{\rm in}|0_{\rm in}\rangle=b_{\bm k}^{\rm in}|0_{\rm in}\rangle=0\,.
\end{eqnarray}
If we denote the vacuum for $\phi_{\bm k}$ by $|0_{\rm in}\rangle_\phi$ and for $\chi_{\bm k}$ by $|0_{\rm in}\rangle_\chi$, then the Bunch-Davies vacuum for the total system is expressed as $|0_{\rm in}\rangle=|0_{\rm in}\rangle_\phi\otimes|0_{\rm in}\rangle_\chi$ where each $|0_{\rm in}\rangle_\phi$ and $|0_{\rm in}\rangle_\chi$ is also the Bunch-Davies vacuum.

Now we consider a state $|\psi\rangle$ defined by Bogoliubov transformations that make a  correlation between the two scalar fields by mixing the operator $a_{\bm k}$ with $b_{\bm k}$,
\begin{eqnarray}
\tilde{a}_{\bm k}=\Gamma_k\,a_{\bm k}^{\rm in}+\Delta_k\,b_{-\bm k}^{\rm in \,\dag}\,,\qquad
\tilde{b}_{\bm k}=\Gamma_k\,b_{\bm k}^{\rm in}+\Delta_k\,a_{-\bm k}^{\rm in \,\dag}\,,
\end{eqnarray}
where $\Gamma_k$ and $\Delta_k$ are Bogoliubov coefficients with $|\Gamma_k|^2-|\Delta_k|^2=1$ and 
\begin{eqnarray}
\tilde{a}_{\bm k}|\psi\rangle=\tilde{b}_{\bm k}|\psi\rangle=0\,.
\end{eqnarray}
This state $|\psi\rangle$ is a non-Bunch-Davies vacuum expressed by a four-mode squeezed state:
\begin{eqnarray}
|\psi\rangle={\tilde N}\exp\left[-\sum_{\bm k}\frac{\Delta_k}{\Gamma_k}
\left(  a_{\bm k}^{\rm in \,\dag} b_{-\bm k}^{\rm in \,\dag} 
+a_{-\bm k}^{\rm in \,\dag}b_{\bm k}^{\rm in \,\dag}  \right) \right]|0_{\rm in}\rangle\,,
\label{four-mode}
\end{eqnarray}
where ${\tilde N}$ is the normalization factor, $|0_{\rm in}\rangle=|0_{\rm in}\rangle_\phi\otimes|0_{\rm in}\rangle_\chi$ and each Bunch-Davies vacuum state $|0_{\rm in}\rangle_\phi$ and $|0_{\rm in}\rangle_\chi$ is written by a two-mode squeezed state
\begin{eqnarray}
|0_{\rm in}\rangle_\phi&\equiv& \prod_{\bm k}|0_{\bm k}^{\rm in}\rangle_\phi 
= \prod_{\bm k}\sum_{n=0}^\infty\frac{\tanh^n r_k}{\cosh r_k}\,
|n_{\bm k}^{\rm out}\rangle_\phi\otimes|n_{-\bm k}^{\rm out}\rangle_\phi\ ,
\nonumber\\
|0_{\rm in}\rangle_\chi&\equiv&\prod_{\bm k}|0_{\bm k}^{\rm in}\rangle_\chi 
=\prod_{\bm k}\sum_{n=0}^\infty\frac{\tanh^n r_k}{\cosh r_k}\,
|n_{\bm k}^{\rm out}\rangle_\chi\otimes|n_{-\bm k}^{\rm out}\rangle_\chi\,.
\label{0-phichi}
\end{eqnarray}

Since the four-mode squeezed state Eq.~(\ref{four-mode}) consists of an infinite sum of states, let us take up to the first order of the Taylor series of Eq.~(\ref{four-mode}) for simplicity, which is expressed as $|\psi\rangle = \prod_{\bm k} |\psi_{\bm k}\rangle$ with
\begin{eqnarray}
|\psi_{\bm k}\rangle=
A_k|0_{\bm k}^{\rm in}\rangle_\phi\otimes|0_{\bm k}^{\rm in}\rangle_\chi
+\frac{B_k}{\sqrt{2}} \bigl(\,
|1_{\bm k}^{\rm in}\rangle_\phi\otimes|1_{-\bm k}^{\rm in}\rangle_\chi 
+|1_{-\bm k}^{\rm in}\rangle_\phi\otimes|1_{\bm k}^{\rm in}\rangle_\chi\,\bigr)\,,
\label{0011}
\end{eqnarray}
where the conservation of probability $|A_k|^2+|B_k|^2=1$ holds. The single particle excitation state is calculated by operating ${\tilde a}_{\bm k}^\dag$ (or ${\tilde b}_{\bm k}^\dag$) on Eq.~(\ref{four-mode}),
\begin{eqnarray}
|1_{\bm k}^{\rm in}\rangle&=&
\sum_{n=0}^\infty\frac{\tanh^n r_k}{\cosh^2 r_k}\sqrt{n+1}\,
|\left(n+1\right)_{\bm k}^{\rm out}\rangle\otimes|n_{-\bm k}^{\rm out}\rangle\,,
\nonumber\\
|1_{-\bm k}^{\rm in}\rangle&=&\sum_{n=0}^\infty\frac{\tanh^n r_k}{\cosh^2 r_k}\sqrt{n+1}\,
|n_{\bm k}^{\rm out}\rangle\otimes|\left(n+1\right)_{-\bm k}^{\rm out}\rangle\,,
\label{1-phichi}
\end{eqnarray}
where we omitted the subscripts $\phi$ or $\chi$ of $|1\rangle$ for simplicity unless there may be any confusion.
Although we truncated the four-mode squeezed state Eq.~(\ref{four-mode}), we can obtain the large enough violation of BMK inequalities as we will see in section~\ref{s4-3}.

\section{Cosmological violation of BMK inequalities}
\label{s4}

According to Eq.~(\ref{maximal1}), we naively expect the violation of the BMK inequalities increases with the number of modes $\bm k$ to measure. However, the upper bound in Eq.~(\ref{maximal1}) is only attained by maximally entangled states. Since the cosmological initial states are not maximally entangled states, in this section, we see how much the Bunch-Davies vacuum and the non-Bunch-Davies vacuum violate the BMK inequalities.

\subsection{Two-mode squeezed state}
\label{s4-1}

Let us check the BMK inequalities for the Bunch-Davies vacuum expressed by a two-mode squeezed state~Eq.~(\ref{two-mode}). Here, we use the pseudospin operators correspond to measuring the parity along various axes in the Hilbert space~\cite{Chen2002}.
The pseudospin operators ${\bm S}$ have eigenvalues $\pm1$ and the inner product with a unit vector ${\bm n}$ is expressed as  
\begin{eqnarray}
\bm n\cdot\bm S=S_z\cos\theta+\sin\theta\left(e^{i\varphi}S_-+e^{-i\varphi}S_+\right)\,,
\label{ns1}
\end{eqnarray}
where the unit vector is chosen as ${\bm n}=\left(\sin\theta\cos\varphi\,,\sin\theta\sin\varphi\,,\cos\theta\right)$ and $\left(\bm n\cdot\bm S\right)^2=I$. 
Since the pseudospin operators act on $|2n+1\rangle$ and $|2n\rangle$ differently, it is convenient to divide the states $n$ into even and odd parity for computation. Focusing on the Hilbert space for a single Fourier mode ${\bm k}$, ${\cal H}_{\bm k}$, Eq.~(\ref{0-phichi}) is written by
\begin{eqnarray}
\hspace{-5mm}|0_{\bm k}^{\rm in}\rangle
&\equiv&\sum_{n=0}^\infty\frac{\tanh^n r_k}{\cosh r_k}\,|n_{\bm k}^{\rm out}\rangle\otimes|n_{-\bm k}^{\rm out}\rangle\nonumber\\
&=&\sum_{n=0}^\infty\frac{\tanh^{2n} r_k}{\cosh r_k}\,|2n_{\bm k}^{\rm out}\rangle\otimes|2n_{-\bm k}^{\rm out}\rangle
+\sum_{n=0}^\infty\frac{\tanh^{2n+1} r_k}{\cosh r_k}\,|\left(2n+1\right)_{\bm k}^{\rm out}\rangle\otimes|\left( 2n+1\right)_{-\bm k}^{\rm out}\rangle\,.
\label{parity}
\end{eqnarray}
For the two-mode squeezed state, we need two sets of non-commuting pseudospin operators as demonstrated in Eq.~(\ref{bell1}). Since we consider two unit vectors for nonprimed operators, we need a plane containing those two vectors. Thus, without any loss of generality, we can take $\varphi=0$ ($x,z$-plane), then Eq.~(\ref{ns1}) is simplified as
\begin{eqnarray}
\bm n\cdot\bm S=S_z\cos\theta+S_x\sin\theta\,.
\end{eqnarray}
By using Eq.~(\ref{B2B3}), the expectation value of Bell operator in the Bunch-Davies vacuum is then written by the psedospin operators as
\begin{eqnarray}
\langle 0_{\bm k}^{\rm in}|{\cal B}_2|0_{\bm k}^{\rm in}\rangle=\frac{1}{2}\left[\,E\left(\theta_1\,,\theta_2\right)+E\left(\theta_1\,,\theta_{2'}\right)+E\left(\theta_{1'}\,,\theta_2\right)
-E\left(\theta_{1'}\,,\theta_{2'}\right)\,\right]\,,
\end{eqnarray}
where ${\cal O}_1\equiv{\bm n}_1\cdot{\bm S}$, ${\cal O}_2\equiv{\bm n}_2\cdot{\bm S}$, ${\cal O}^\prime_1\equiv{\bm n}^\prime_1\cdot{\bm S}$, ${\cal O}^\prime_2\equiv{\bm n}^\prime_2\cdot{\bm S}$ in Eq.~(\ref{B2B3}). And $E\left(\theta_1\,,\theta_2\right)$ is 
\begin{eqnarray}
E\left(\theta_1\,,\theta_2\right)&=&\langle 0_{\bm k}^{\rm in}|\left(S_z\cos\theta_1+S_x\sin\theta_1\right)\otimes
\left(S_z\cos\theta_2+S_x\sin\theta_2
\right)|0_{\bm k}^{\rm in}\rangle\,,\nonumber\\
&=&\cos\theta_1\cos\theta_2+\tanh 2r_k\,\sin\theta_1\sin\theta_2\,.
\end{eqnarray}
Here, we used Eqs.~(\ref{rules}) and (\ref{parity}). Choosing $\theta_1=0\,,\theta_{1'}=\pi/2\,,\theta_2=-\theta_{2'}$
we get
\begin{eqnarray}
\langle 0_{\bm k}^{\rm in}|{\cal B}_2|0_{\bm k}^{\rm in}\rangle=\cos\theta_2+\tanh 2r_k\,\sin\theta_2\,.
\end{eqnarray}
For $\theta_2=\tan^{-1}\tanh 2r_k$, we get the maximal violation which has the extra $\sqrt{2}$ factor: 
\begin{eqnarray}
\langle 0_{\bm k}^{\rm in}|{\cal B}_2|0_{\bm k}^{\rm in}\rangle
=\sqrt{1+\tanh^2 2r_k}\leq\sqrt{2} \,. 
\end{eqnarray}
where the maximal value is obtained in the infinite squeezing limit $r_k\rightarrow\infty$.
We reproduced Bell inequality of a pair of spins in section~\ref{s2-1} by using continuous quantum variables~\cite{Chen2002} and found that the Bell inequality is maximally violated by the Bunch-Davies vacuum according to Eq.~(\ref{maximal1}) with $n=2$.
As we see in section.\ref{s2-3}, this two-mode squeezed state corresponds to the case of $N=2$, $L=1$, $K_1=0$. Thus we get $\langle{\cal B}_2\rangle^2+\langle{\cal B}'_2\rangle^2\leq 2$. Although we focused on a single Fourier mode ${\bm k}$ (a pair of spins or a two-partite system), the Bunch-Davies vacuum consists of infinite products of ${\bm k}$ as in Eq.~(\ref{two-mode}).  
So, if we increase the number of modes to measure, say $m$-pairs $(m=2,3,4\cdots)$, it appears the violation increases by Eq.~(\ref{maximal1}) with $n=2m$. However, this case corresponds to $N=2m$, $L=m$, $K_1=0$, which holds $N-2L=0$ in the classification Eq.~(\ref{E}). Thus, the violation of BMK inequalities does not increase anymore.

\subsection{Four-mode squeezed state}
\label{s4-2}

Let us see the BMK inequalities for a non-Bunch-Davies vacuum expressed by the four-mode squeezed state Eq.~(\ref{0011}) next. In this case, we can expect the violation increases with the number of modes $\bm k$ to measure because the four-mode squeezed state realizes $N=4m$, $L=m$, $K_1=0$ and then $N-2L\neq 0$ in the classification Eq.~(\ref{E}).

\begin{figure}[t]
\begin{center}
\vspace{-2cm}
\includegraphics[height=9.5cm]{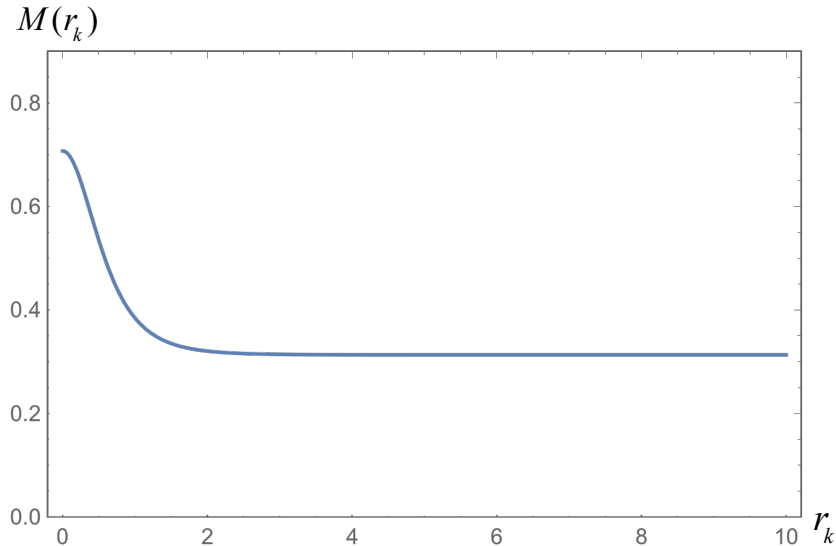}\centering
\vspace{-1.5cm}
\caption{Plot of the behavior of quantum interference as a function of $r_k$. 
The quantum interference remains finite in inflation. The asymptotic value is about $0.3$.}
\label{fig1}
\end{center}
\end{figure}

To obtain the Bell operator ${\cal B}_4$, we use the Mermin-Klyshiko operator Eq.~(\ref{MK1}) recursively to find\footnote{We cannot use Eq.~(\ref{MK2}) with $n=4$, $p=2$ to calculate the expectation value of ${\cal B}_4$ in the four-mode squeezed state because in this formula, the expectation values of ${\cal B}_{n-p}$ and ${\cal B}_p$ are supposed to be unentangled. Thus we use Eq.~(\ref{B4}) which is derived recursively by using Eq.~(\ref{MK1}).}
\begin{eqnarray}
4{\cal B}_4&=&-{\cal O}_1\otimes{\cal O}_2\otimes{\cal O}_3\otimes{\cal O}_4
-{\cal O}'_1\otimes{\cal O}'_2\otimes{\cal O}'_3\otimes{\cal O}'_4
+{\cal O}_1\otimes{\cal O}_2\otimes{\cal O}_3\otimes{\cal O}'_4\nonumber\\
&&
+{\cal O}_1\otimes{\cal O}_2\otimes{\cal O}'_3\otimes{\cal O}_4
+{\cal O}_1\otimes{\cal O}'_2\otimes{\cal O}_3\otimes{\cal O}_4
+{\cal O}'_1\otimes{\cal O}_2\otimes{\cal O}_3\otimes{\cal O}_4\nonumber\\
&&
+{\cal O}_1\otimes{\cal O}_2\otimes{\cal O}'_3\otimes{\cal O}'_4
+{\cal O}_1\otimes{\cal O}'_2\otimes{\cal O}_3\otimes{\cal O}'_4
+{\cal O}'_1\otimes{\cal O}_2\otimes{\cal O}_3\otimes{\cal O}'_4\nonumber\\
&&
+{\cal O}_1\otimes{\cal O}'_2\otimes{\cal O}'_3\otimes{\cal O}_4
+{\cal O}'_1\otimes{\cal O}_2\otimes{\cal O}'_3\otimes{\cal O}_4
+{\cal O}'_1\otimes{\cal O}'_2\otimes{\cal O}_3\otimes{\cal O}_4\nonumber\\
&&
-{\cal O}_1\otimes{\cal O}'_2\otimes{\cal O}'_3\otimes{\cal O}'_4
-{\cal O}'_1\otimes{\cal O}_2\otimes{\cal O}'_3\otimes{\cal O}'_4
-{\cal O}'_1\otimes{\cal O}'_2\otimes{\cal O}_3\otimes{\cal O}'_4\nonumber\\
&&-{\cal O}'_1\otimes{\cal O}'_2\otimes{\cal O}'_3\otimes{\cal O}_4\,,
\label{B4}
\end{eqnarray}
where ${\cal O}_1\equiv n_1\cdot\bm S$, ${\cal O}_2\equiv n_2\cdot\bm S$, ${\cal O}_3\equiv n_3\cdot\bm S$ and ${\cal O}_4\equiv n_4\cdot\bm S$ etc. The expectation value of the above first term in the non-Bunch-Davies vacuum is computed as follows. 
We calculate the expectation value of the following operator
\begin{eqnarray}
{\cal E}_4\equiv\left(\bm n_1\cdot\bm S\right)\otimes\left(\bm n_2\cdot\bm S\right)\otimes\left(\bm n_3\cdot\bm S\right)\otimes\left(\bm n_4\cdot\bm S\right) \,.
\end{eqnarray}
Focusing on a single mode ${\bm k}$ of Eq.~(\ref{0011}), the expectation value of the above operator in the non-Bunch-Davies vacuum is written by
\begin{eqnarray}
\langle\psi_{\bm k}|{\cal E}_4|\psi_{\bm k}\rangle
&=&|A_k|^2{}_\phi\langle 0_{\bm k}^{\rm in}|\otimes{}_\chi\langle 0_{\bm k}^{\rm in}|{\cal E}_4|0_{\bm k}^{\rm in}\rangle_\phi\otimes|0_{\bm k}^{\rm in}\rangle_\chi \nonumber\\
&&+\frac{A^*_k B_k}{\sqrt{2}}\,{}_\phi\langle 0_{\bm k}^{\rm in}|\otimes{}_\chi\langle 0_{\bm k}^{\rm in}|{\cal E}_4\,\Bigl(\,|1_{\bm k}^{\rm in}\rangle_\phi\otimes|1_{-\bm k}^{\rm in}\rangle_\chi 
+ |1_{-\bm k}^{\rm in}\rangle_\phi\otimes|1_{\bm k}^{\rm in}\rangle_\chi\,\Bigr) \nonumber\\
&&+\frac{A_k B^*_k}{\sqrt{2}}\,
\Bigl(\,{}_\phi\langle 1_{\bm k}^{\rm in}|\otimes{}_\chi\langle 1_{-\bm k}^{\rm in}| 
+{}_\phi\langle 1_{-\bm k}^{\rm in}|\otimes{}_\chi\langle 1_{\bm k}^{\rm in}|\,\Bigr)
{\cal E}_4|0_{\bm k}^{\rm in}\rangle_\phi\otimes|0_{\bm k}^{\rm in}\rangle_\chi \nonumber\\
&& +\frac{|B_k|^2}{2}\,
\Bigl(\,{}_\phi\langle 1_{\bm k}^{\rm in}|\otimes{}_\chi\langle 1_{-\bm k}^{\rm in}| 
+{}_\phi\langle 1_{-\bm k}^{\rm in}|\otimes{}_\chi\langle 1_{\bm k}^{\rm in}|\,\Bigr)
{\cal E}_4\,
\Bigl(\,|1_{\bm k}^{\rm in}\rangle_\phi\otimes|1_{-\bm k}^{\rm in}\rangle_\chi 
+|1_{-\bm k}^{\rm in}\rangle_\phi\otimes|1_{\bm k}^{\rm in}\rangle_\chi\,\Bigr)
\nonumber \\
&\equiv& E(\theta_1,\,\theta_2,\,\theta_3,\,\theta_4,\,\varphi_1,\,\varphi_2,\,\varphi_3,\,\varphi_4 )\,,
\label{E4}
\end{eqnarray}
where the states $|0_{\bm k}\rangle$ and $|1_{\bm k}\rangle$ are given in Eqs.~(\ref{0-phichi}) and (\ref{1-phichi}) respectively. 
Note that each term can be factorized into a product of ${}_\phi\langle I | \left(\bm n_1\cdot\bm S\right)\otimes\left(\bm n_2\cdot\bm S\right)|J\rangle_\phi$ and ${}_\chi\langle I| \left(\bm n_3\cdot\bm S\right)\otimes\left(\bm n_4\cdot\bm S\right)|J \rangle_\chi$, where $I,J=0_{{\bm k}},1_{\pm\bm k}$.
By using Eqs.~(\ref{rules}) and (\ref{parity}), we find
\begin{eqnarray}
&&\hspace{-0.5cm}E\left(\theta_1,\theta_2,\theta_3,\theta_4,\varphi_1,\varphi_2,\varphi_3,\varphi_4 \right)\nonumber\\
&& =
|A_k|^2f\left(\theta_1,\theta_2,\varphi_1,\varphi_2\right)
f\left(\theta_3,\theta_4,\varphi_3,\varphi_4\right)               \\
&& +
\frac{A^*_k B_k}{\sqrt{2}}\,\Bigl[ g_+ \left(\theta_1,\theta_2,\varphi_1,\varphi_2\right)
g_- \left(\theta_3,\theta_4,\varphi_3,\varphi_4\right) 
          + g_- \left(\theta_1,\theta_2,\varphi_1,\varphi_2\right)
g_+ \left(\theta_3,\theta_4,\varphi_3,\varphi_4\right) \Bigr]\nonumber\\
&& + \frac{A_k B^*_k}{\sqrt{2}}\, \Bigl[ g_+^*\left(\theta_1,\theta_2,\varphi_1,\varphi_2\right)
g_-^*\left(\theta_3,\theta_4,\varphi_3,\varphi_4\right) + g_-^*\left(\theta_1,\theta_2,\varphi_1,\varphi_2\right)
g_+^*\left(\theta_3,\theta_4,\varphi_3,\varphi_4\right)\Bigr]\nonumber\\
&&+\frac{|B_k|^2}{2}\, \Bigl[
       h_{++}\left(\theta_1,\theta_2,\varphi_1,\varphi_2\right) h_{--}\left(\theta_3,\theta_4,\varphi_3,\varphi_4\right)
       +  h_{--}\left(\theta_1,\theta_2,\varphi_1,\varphi_2\right) h_{++}\left(\theta_3,\theta_4,\varphi_3,\varphi_4\right) \Bigr.\nonumber\\
&& \Bigl.  \qquad \qquad  +  h_{+-}\left(\theta_1,\theta_2,\varphi_1,\varphi_2\right) h_{-+}\left(\theta_3,\theta_4,\varphi_3,\varphi_4\right)
       +  h_{-+}\left(\theta_1,\theta_2,\varphi_1,\varphi_2\right) h_{+-}\left(\theta_3,\theta_4,\varphi_3,\varphi_4\right)
 \Bigr] \nonumber \,,
\end{eqnarray}
where  we defined
\begin{eqnarray}
f\left(\theta_1,\theta_2 ,\varphi_1,\varphi_2\right)&=&
 {}_\phi\langle 0_{\bm k}^{\rm in}|\left(\bm n_1\cdot\bm S\right)\otimes\left(\bm n_2\cdot\bm S\right)|0_{\bm k}^{\rm in}\rangle_\phi\nonumber\\
&=&\cos\theta_1\cos\theta_2+\tanh 2r_k\,\cos\left(\varphi_1+\varphi_2\right)\sin\theta_1\sin\theta_2 \,,
\nonumber\\
g_+ \left(\theta_1,\theta_2 ,\varphi_1,\varphi_2\right) 
&=& {}_\phi\langle 0_{\bm k}^{\rm in}|\left(\bm n_1\cdot\bm S\right)\otimes\left(\bm n_2\cdot\bm S\right)|1_{\bm k}^{\rm in}\rangle_\phi\nonumber\\
&=&\left(-e^{i\varphi_1}\sin\theta_1\cos\theta_2+\tanh r_k\,
e^{-i\varphi_2}\cos\theta_1\sin\theta_2\right)M(r_k)\,,
\nonumber\\
g_- \left(\theta_1,\theta_2 ,\varphi_1,\varphi_2\right) 
&=& {}_\phi\langle 0_{\bm k}^{\rm in}|\left(\bm n_1\cdot\bm S\right)\otimes\left(\bm n_2\cdot\bm S\right)|1_{-\bm k}^{\rm in}\rangle_\phi\nonumber\\
&=&\left(-e^{i\varphi_2}\cos\theta_1\sin\theta_2  +\tanh r_k \  e^{-i\varphi_1}\sin\theta_1\cos\theta_2\right)M(r_k) \,,
\nonumber\\
h_{++}\left(\theta_1,\theta_2 ,\varphi_1 ,\varphi_2\right) 
&=& {}_\phi\langle 1_{\bm k}^{\rm in}|\left(\bm n_1\cdot\bm S\right)\otimes\left(\bm n_2\cdot\bm S\right)|1_{\bm k}^{\rm in}\rangle_\phi\nonumber\\
&=&   -\cos\theta_1\cos\theta_2\nonumber\\
&=& h_{--}\left(\theta_1,\theta_2 ,\varphi_1 ,\varphi_2\right)\,,\nonumber\\
h_{+-}\left(\theta_1,\theta_2 ,\varphi_1 ,\varphi_2\right)
&=& {}_\phi\langle 1_{\bm k}^{\rm in}|\left(\bm n_1\cdot\bm S\right)\otimes\left(\bm n_2\cdot\bm S\right)|1_{-\bm k}^{\rm in}\rangle_\phi\nonumber\\
&=&\frac{1+\tanh^4r_k}{\left(1+\tanh^2r_k\right)^2}
\exp\left(-i \varphi_1 +i \varphi_2\right)  \sin\theta_1\sin\theta_2\nonumber\\
&=& h_{-+}\left(\theta_1,\theta_2 ,\varphi_1 ,\varphi_2\right)^* 
\,,
\end{eqnarray}
and $f\left(\theta_3,\theta_4 ,\varphi_3,\varphi_4\right)$, $g_\pm \left(\theta_3,\theta_4,\varphi_3,\varphi_4\right)$, $h_{\pm\mp}\left(\theta_3,\theta_4,\varphi_3,\varphi_4\right)$ and $h_{\pm\pm}\left(\theta_3,\theta_4,\varphi_3,\varphi_4\right)$ are obtained by interchanging $\theta_1,\theta_2 ,\varphi_1,\varphi_2$ with $\theta_3,\theta_3 ,\varphi_3,\varphi_4$, respectively. We also defined
\begin{eqnarray}
M(r_k) = \sum_{n=0}^\infty\frac{\tanh^{4n}r_k}{\cosh^3r_k}\sqrt{\frac{2n+1}{2}}=\Phi\left(\tanh^4r_k,\,-\frac{1}{2},\,\frac{1}{2}\right)\frac{1}{\cosh^3r_k}\,,
\end{eqnarray}
where $\Phi\left(\tanh^4r_k,\,-\frac{1}{2},\,\frac{1}{2}\right)$ 
is the Lerch transcendent. 
Note that $M(r_k)$ appears only in the off-diagonal elements which represent quantum interference. We find that $M(r_k)$ converges to a finite value 
 in the large squeezed limit $r_k\rightarrow \infty$ as shown in Figure~\ref{fig1}.

Thus the expectation value of the Mermin-Klyshko operator ${\cal B}_4$ in Eq.~(\ref{B4}) is given by
\begin{eqnarray}
\hspace{-0.5cm}\langle\psi_{\bm k}|{\cal B}_4|\psi_{\bm k}\rangle 
&=&\frac{1}{4}\Bigl[ 
-E\left(\theta_1,\theta_2,\theta_3,\theta_4,\varphi_1,\varphi_2, \varphi_3, \varphi_4\right) 
-E\left(\theta_1^\prime,\theta_2^\prime,\theta_3^\prime,\theta_4^\prime,\varphi_1^\prime,\varphi_2^\prime, \varphi_3^\prime , \varphi_4^\prime\right) 
\Bigr.\nonumber\\
&&\quad+E\left(\theta_1,\theta_2,\theta_3,\theta_4^\prime,\varphi_1,\varphi_2, \varphi_3, \varphi_4^\prime\right)
+E\left(\theta_1,\theta_2,\theta_3^\prime,\theta_4,\varphi_1,\varphi_2, \varphi_3^\prime, \varphi_4\right)
\nonumber\\
&&\quad
+E\left(\theta_1,\theta_2^\prime,\theta_3,\theta_4,\varphi_1,\varphi_2^\prime, \varphi_3, \varphi_4\right)
+E\left(\theta_1^\prime,\theta_2,\theta_3,\theta_4,\varphi_1^\prime,\varphi_2, \varphi_3, \varphi_4\right)
\nonumber\\
&&\quad
+E\left(\theta_1,\theta_2,\theta_3^\prime,\theta_4^\prime,\varphi_1,\varphi_2, \varphi_3^\prime, \varphi_4^\prime\right)
+E\left(\theta_1,\theta_2^\prime,\theta_3,\theta_4^\prime,\varphi_1,\varphi_2^\prime, \varphi_3, \varphi_4^\prime\right)
\nonumber\\
&&\quad
+E\left(\theta_1^\prime,\theta_2,\theta_3,\theta_4^\prime,\varphi_1^\prime,\varphi_2, \varphi_3, \varphi_4^\prime\right)
+E\left(\theta_1,\theta_2^\prime,\theta_3^\prime,\theta_4,\varphi_1,\varphi_2^\prime, \varphi_3^\prime, \varphi_4\right)
\nonumber\\
&&\quad
+E\left(\theta_1^\prime,\theta_2,\theta_3^\prime,\theta_4,\varphi_1^\prime,\varphi_2, \varphi_3^\prime, \varphi_4\right)
+E\left(\theta_1^\prime,\theta_2^\prime,\theta_3,\theta_4,\varphi_1^\prime,\varphi_2^\prime, \varphi_3, \varphi_4\right)
\nonumber\\
&&\quad
-E\left(\theta_1,\theta_2^\prime,\theta_3^\prime,\theta_4^\prime,\varphi_1,\varphi_2^\prime, \varphi_3^\prime, \varphi_4^\prime\right)
-E\left(\theta_1^\prime,\theta_2,\theta_3^\prime,\theta_4^\prime,\varphi_1^\prime,\varphi_2, \varphi_3^\prime, \varphi_4^\prime\right)
\nonumber\\
&&\quad\Bigl.
-E\left(\theta_1^\prime,\theta_2^\prime,\theta_3,\theta_4^\prime,\varphi_1^\prime,\varphi_2^\prime, \varphi_3, \varphi_4^\prime\right)
-E\left(\theta_1^\prime,\theta_2^\prime,\theta_3^\prime,\theta_4,\varphi_1^\prime,\varphi_2^\prime, \varphi_3^\prime, \varphi_4\right)\Bigr]\,.
\end{eqnarray}

\begin{figure}[t]
\begin{center}
\vspace{-2cm}
\includegraphics[height=9.5cm]{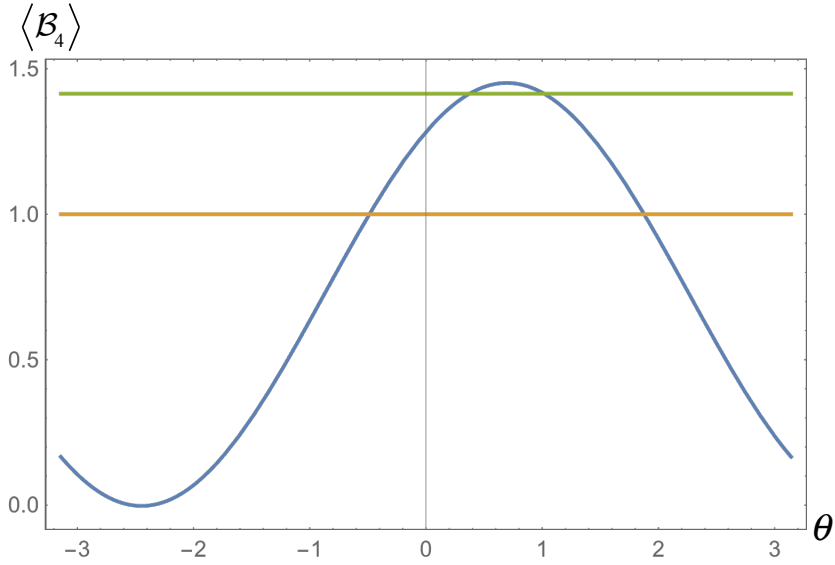}
\vspace{-1.5cm}
\caption{Plot of the violation of the BMK inequalities. The blue line is for ${\cal B}_4$. $A_k$ and $B_k$ have been set to $\sqrt{0.95}$ and $\sqrt{0.05}$ and $r_k=1.7$. The orange line is the classical upper bound and the green line is $\sqrt{2}$ which is the quantum upper bound for the Bunch-Davies vacuum. The part exceeding the green line grows exponentially as the number of modes to measure increases according to Eq.~(\ref{increase}). Note that the plot is parametrized by only one parameter $\theta$.}
\label{fig2}
\end{center}
\end{figure}

Now we consider higher order BMK inequaliries by increasing the number of modes to measure. We compute the expectation value of ${\cal B}_8$ in the Hilbert spaces ${\cal H}_{{\bm k}_{1}}\otimes{\cal H}_{{\bm k}_{2}}$ by using Eq.~(\ref{812}) as follows
\begin{eqnarray}
\langle\psi_{{\bm k}_{1}} \psi_{{\bm k}_{2}}|{\cal B}_8|\psi_{{\bm k}_{1}} \psi_{{\bm k}_{2}}\rangle  
&=&\frac{1}{2}\,\langle\psi_{{\bm k}_{1}}|{\cal B}_4|\psi_{{\bm k}_{1}}\rangle  
\Bigl(\langle\psi_{{\bm k}_{2}}|{\cal B}_4|\psi_{{\bm k}_{2}}\rangle 
+\langle\psi_{{\bm k}_{2}}|{\cal B}'_4|\psi_{{\bm k}_{2}}\rangle \Bigr)  \nonumber \\
&&+\frac{1}{2}\,\langle\psi_{{\bm k}_{1}}|{\cal B}'_4|\psi_{{\bm k}_{1}}\rangle 
\Bigl(\langle\psi_{{\bm k}_{2}}|{\cal B}_4|\psi_{{\bm k}_{2}}\rangle 
-\langle\psi_{{\bm k}_{2}}|{\cal B}'_4|\psi_{{\bm k}_{2}}\rangle \Bigr)\,,
\label{EXP8}
\end{eqnarray}
where for notational convenience we have defined $|\psi_{{\bm k}_{1}} \psi_{{\bm k}_{2}}\rangle=|\psi_{{\bm k}_{1}}\rangle\otimes|\psi_{{\bm k}_{2}}\rangle$ and we assumed  there is no correlation between different groups ${\cal B}_{n-p}$ and ${\cal B}_p$.
Using the above result, we can further calculate the expectation value of ${\cal B}_{12}$ in the Hilbert spaces ${\cal H}_{{\bm k}_{1}}\otimes{\cal H}_{{\bm k}_{2}}\otimes {\cal H}_{{\bm k}_{3}}$ as
\begin{eqnarray}
\hspace{-10mm}
\langle\psi_{{\bm k}_{1}} \psi_{{\bm k}_{2}}\psi_{{\bm k}_{3}}|{\cal B}_{12}|\psi_{{\bm k}_{1}} \psi_{{\bm k}_{2}}\psi_{{\bm k}_{3}}\rangle 
&=&\frac{1}{2}\,\langle\psi_{{\bm k}_{1}} \psi_{{\bm k}_{2}}|{\cal B}_8|\psi_{{\bm k}_{1}} \psi_{{\bm k}_{2}}\rangle 
\Bigl(\langle\psi_{{\bm k}_{3}}|{\cal B}_4|\psi_{{\bm k}_{3}}\rangle 
+\langle\psi_{{\bm k}_{3}}|{\cal B}'_4|\psi_{{\bm k}_{3}}\rangle \Bigr)  \nonumber \\
&&+\frac{1}{2}\,\langle\psi_{{\bm k}_{1}} \psi_{{\bm k}_{2}}|{\cal B}'_8|\psi_{{\bm k}_{1}} \psi_{{\bm k}_{2}}\rangle 
\Bigl(\langle\psi_{{\bm k}_{3}}|{\cal B}_4|\psi_{{\bm k}_{3}}\rangle 
 -\langle\psi_{{\bm k}_{3}}|{\cal B}'_4|\psi_{{\bm k}_{3}}\rangle \Bigr)\,,
\label{EXP12}
\end{eqnarray}
We plotted the expectation value of Mermin-Klyshko oeprator ${\cal B}_4$ in Figure~\ref{fig2} where we see that the violation of the BMK inequalities exceeds the quantum upper bound for the Bunch-Davies vacuum. In our truncated four-mode squeezed state Eq.~(\ref{0011}), we found the maximum value of ${\cal B}_4$
is $1.45$ with $A_k=\sqrt{0.95}, B_k=\sqrt{0.05}$, and $r_k=1.7$. The violation of of the BMK inequalities of ${\cal B}_8$ in Eq.~(\ref{EXP8}) and ${\cal B}_{12}$ in Eq.~(\ref{EXP12}) become bigger as explained in the next subsection. This is because the four-mode squeezed state holds $N-2L\neq 0$ in the classification Eq.~(\ref{E}).

\subsection{Infinite violation of BMK inequalities}
\label{s4-3}

In the previous subsection, we first focused on the Hilbert space for a single Fourier mode ${\bm k}$, ${\cal H}_{\bm k}$,
and extended the analysis to ${\cal H}_{{\bm k}_{1}}\otimes{\cal H}_{{\bm k}_{2}}$ and ${\cal H}_{{\bm k}_{1}}\otimes{\cal H}_{{\bm k}_{2}}\otimes {\cal H}_{{\bm k}_{3}}$. 
 However, the four-mode squeezed state consists of infinite products of ${\bm k}$ as in Eq.~(\ref{0-phichi}). 
 Let's see the upper bound of the quadratic form of Mermin-Klyshko inequalities when we increase the number of modes ${\bm k}$ to measure. 
 
If we plug the Mermin-Klyshko operators Eq.~(\ref{MK2}) into the quadratic form of Bell inequality Eq.~(\ref{quadratic}), we obtain
\begin{eqnarray}
{\cal M}_N&=&\langle{\cal B}_N\rangle^2+\langle{\cal B}_N^\prime\rangle^2\nonumber\\
&=&\frac{1}{2}\left(\langle{\cal B}_{N-p}\rangle^2+\langle{\cal B}_{N-p}^\prime\rangle^2\right)
\left(\langle{\cal B}_{p}\rangle^2+\langle{\cal B}_{p}^\prime\rangle^2\right)\nonumber\\
&=&\frac{1}{2}{\cal M}_{N-p}\,{\cal M}_p\,,
\label{recursive}
\end{eqnarray}
where we assumed that there is no correlation between ${\cal B}_{N}$ and ${\cal B}_{N-p}$, that is, $\langle{\cal B}_N{\cal B}_{N-p}\rangle^2=\langle{\cal B}_N\rangle^2\langle{\cal B}_{N-p}\rangle^2$.

For a four-mode squeezed state, we take $N=4n~(n=1,2,3\cdots)$ where $n$ corresponds to the number of modes ${\bm k}$ to measure and $p=4$, then we have
\begin{eqnarray}
{\cal M}_{4n}=\frac{1}{2}{\cal M}_{4n-4}\,{\cal M}_4=\left(\frac{1}{2}\right)^{n-1}{\cal M}_{4n-4(n-1)}\,{\cal M}_4^{n-1}=\left(\frac{1}{2}\right)^{n-1}{\cal M}_4^n\,,
\end{eqnarray}
where we used the relation Eq.~(\ref{recursive}) recursively. If we write the maximal violation of ${\cal M}_4$ by $q$, we have
\begin{eqnarray}
{\cal M}_{4n} = \left(\frac{1}{2}\right)^{n-1}q^n=2^{\left(\log_2q-1\right)n+1}\,,
\label{increase}
\end{eqnarray}
then we see that the violation increases exponentially as $n$ increases when $\log_2q-1>1$. In our case, we get the maximum value $\langle{\cal B}_4\rangle\sim 1.45$ for $A_k=\sqrt{0.95}$, $B_k=\sqrt{0.05}$ and $r_k=1.7$, then $q\geq(1.45)^2\simeq 2.1$ and then $\log_2 2.1 \simeq 1.07 >1$. Thus, we have shown that the violation of BMK inequalities increases exponentially with the number of modes to measure $n$. Note that for the two-mode squeezed state, we get the same inequality as above and find the fixed upper bound ${\cal M}_{2n}=2$ because of $q=(\sqrt{2})^2$.

\section{Summary and discussion}
\label{s5}

We studied the violation of the BMK inequalities in initial quantum states of scalar fields in inflation. We showed that the Bell inequality is maximally violated by the Bunch-Davies vacuum which is a two-mode squeezed state of a scalar field. However, it is found that the violation of the BMK inequalities does not increase anymore with the number of modes to measure. We then considered a non-Bunch-Davies vacuum expressed by a four-mode squeezed state of two scalar fields. Remarkably, we found  
that the violation increases exponentially with the number of modes to measure. This result indicates that some evidence that our universe has a quantum mechanical origin may survive in CMB data even if quantum entanglement decays exponentially afterward due to decoherence in the course of evolution of the universe. 

We truncated the four-mode squeezed state Eq.~(\ref{four-mode}) and took the form of Eq.~(\ref{0011}) for simplicity, but we obtained the large enough violation of the BMK inequalities. We expect the larger violation of the BMK inequalities if we use the full form of Eq.~(\ref{four-mode}).

Since we found a clear difference in the violation of the BMK inequalities between the Bunch-Davies vacuum and a non-Bunch-Davies vacuum, we may be able to find the nature of the initial state of the universe. 
The four-mode squeezed state can also be realized if our universe is entangled with another universe initially~\cite{Kanno:2015lja}. Thus, we may be able to test the existence of the other universes by using the difference in the violation of the BMK inequalities.

In this paper, we focused only on the two-mode and the four-mode squeezed states, but it is easy to investigate the other type of squeezed states
 in a similar way. As far as the relation $N-2L\neq 0$ in Eq.~(\ref{E}) holds, we can expect the large enough violation of the BMK inequalities. The BMK inequalities in a multipartite system are known to be maximally violated by the Greenberger-Horne-Zeilinger (GHZ) state~\cite{GHZ}, so it would be interesting to discuss the GHZ state in the context of cosmology.

Although some signatures of a quantum mechanical origin may remain in CMB data, we have not yet found a way to distinguish them from classical density fluctuations. In that case, we might need to find some appropriate cosmological observables as Maldacena considered to perform a Bell type experiment during inflation~\cite{Maldacena:2015bha}. 
It would be interesting to examine Maldacena's model by using the BMK inequalities as a multipartite system.

Gravitational-wave astronomy opens up a new window to explore the universe, so it would be of interest to investigate the BMK inequalities by using gravitational waves. In fact, scalar-tensor initial state entanglement is discussed in~\cite{Collins:2016ahj,Bolis:2016vas}. To this end, we need to come up with an appropriate cosmological observables associated with BMK operators.

\section*{Acknowledgments}
SK was supported by IKERBASQUE, the Basque Foundation 
for Science and the Basque Government (IT-979-16),  
and Spanish Ministry MINECO  (FPA2015-64041-C2-1P). 
JS was supported by JSPS KAKENHI Grant Number 17H02894 and MEXT KAKENHI Grant Number 15H05895.


\begin{thebibliography}{99}


%\cite{Einstein:1935rr}
\bibitem{Einstein:1935rr} 
  A.~Einstein, B.~Podolsky and N.~Rosen,
  %``Can quantum mechanical description of physical reality be considered complete?,''
  Phys.\ Rev.\  {\bf 47}, 777 (1935).
%  doi:10.1103/PhysRev.47.777
  %%CITATION = doi:10.1103/PhysRev.47.777;%%

%\cite{Bell:1964kc}
\bibitem{Bell:1964kc} 
  J.~S.~Bell,
  %``On the Einstein-Podolsky-Rosen paradox,''
  Physics {\bf 1}, 195 (1964).
  %%CITATION = PYCSA,1,195;%%

%\cite{Clauser:1969ny}
\bibitem{Clauser:1969ny} 
  J.~F.~Clauser, M.~A.~Horne, A.~Shimony and R.~A.~Holt,
  %``Proposed experiment to test local hidden variable theories,''
  Phys.\ Rev.\ Lett.\  {\bf 23}, 880 (1969).
 % doi:10.1103/PhysRevLett.23.880
  %%CITATION = doi:10.1103/PhysRevLett.23.880;%%
      
%\cite{Aspect:1981zz}
\bibitem{Aspect:1981zz} 
  A.~Aspect, P.~Grangier and G.~Roger,
  %``Experimental Tests of Realistic Local Theories via Bell's Theorem,''
  Phys.\ Rev.\ Lett.\  {\bf 47}, 460 (1981).\\
%  doi:10.1103/PhysRevLett.47.460
  %%CITATION = doi:10.1103/PhysRevLett.47.460;%%
   A.~Aspect, J.~Dalibard and G.~Roger,
  %``Experimental test of Bell's inequalities using time varying analyzers,''
  Phys.\ Rev.\ Lett.\  {\bf 49}, 1804 (1982).
%  doi:10.1103/PhysRevLett.49.1804
  %%CITATION = doi:10.1103/PhysRevLett.49.1804;%%

%\cite{loop}
\bibitem{Loop}
B. Hensen {\it et al.} 
%"Loophole-free Bell inequality violation using electron spins separated by 1.3 kilometres." 
Nature {\bf 526}, 682-686 (2015).
%doi:10.1038/nature15759
[arXiv:1508.05949 [quant-ph]].\\
M. Giustina {\it et al.} 
%"Significant-loophole-free test of Bell’s theorem with entangled photons." 
Phys.\ Rev.\ Lett. {\bf 115}, 250401 (2015).
%doi:10.1103/PhysRevLett.115.250401
[arXiv:1511.03190 [quant-ph]].\\
L.~K.~Shalm {\it et al.}  
%"Strong loophole-free test of local realism." 
Phys.\ Rev.\ Lett. {\bf 115}, 250402 (2015).
%doi:10.1103/PhysRevLett.115.250402
[arXiv:1511.03189 [quant-ph]].

%\cite{Cirelson:1980ry}
\bibitem{Cirelson:1980ry} 
  B.~S.~Cirelson,
  %``Quantum Generalizations Of Bell's Inequality,''
  Lett.\ Math.\ Phys.\  {\bf 4}, 93 (1980).
%  doi:10.1007/BF00417500
  %%CITATION = doi:10.1007/BF00417500;%%

%\cite{Mermin}
\bibitem{Mermin}
N.~D.~Mermin,
Phys.\ Rev.\ Lett.\ {\bf 65}, 1838  (1990).
%doi:10.1103/PhysRevLett.65.1838

%\cite{B_K}
\bibitem{B_K}
A. V. Belinski and D. N. Klyshko,
%Interference of light and Bell's theorem,
 Physics-Uspekhi,  {\bf 36}, 653 (1993).

%\cite{Gisin:1998ze}
\bibitem{Gisin:1998ze} 
  N.~Gisin and H.~Bechmann-Pasquinucci,
  %``Bell inequality, Bell states and maximally entangled states for n qubits,''
  Phys.\ Lett.\ A {\bf 246}, 1 (1998)
  %doi:10.1016/S0375-9601(98)00516-7
  [quant-ph/9804045].
  %%CITATION = doi:10.1016/S0375-9601(98)00516-7;%%

%\cite{Werner}
\bibitem{Werner}
  R.~F.~Werner, M.~M.~Wolf,
  %``Bell's inequalities for states with positive partial transpose"
 Phys.\ Rev.\ A {\bf 61}, 062102 (2000)
 %doi:10.1103/PhysRevA.61.062102
[arXiv:quant-ph/9910063].
  
%\cite{Alsina}
\bibitem{Alsina}
  D.~Alsina, , A.~Cervera, D.~Goyeneche, J.~I.~Latorre, and K.~\ifmmode \dot{Z}\else \.{Z}\fi{}yczkowski,
%``Operational approach to Bell inequalities: Application to qutrits"
 Phys.\ Rev.\ A {\bf 94}, 032102 (2016)
%doi:10.1103/PhysRevA.94.032102
[arXiv:1606.01991 [quant-ph]].

%\cite{Chen2002}
\bibitem{Chen2002} 
Z.~B.~Chen, J.~W.~Pan, G.~Hou, and Y.~D.~Zhang,
Phys.\ Rev.\ Lett.\  {\bf88}, 040406 (2002).
% doi:10.1103/PhysRevLett.88.040406

%\cite{Martin:2016tbd}
\bibitem{Martin:2016tbd} 
  J.~Martin and V.~Vennin,
  %``Bell inequalities for continuous-variable systems in generic squeezed states,''
  Phys.\ Rev.\ A {\bf 93}, no. 6, 062117 (2016)
%  doi:10.1103/PhysRevA.93.062117
  [arXiv:1605.02944 [quant-ph]].
  %%CITATION = doi:10.1103/PhysRevA.93.062117;%%

%\cite{Nagata}
\bibitem{Nagata}
 K.~Nagata, and M.~Koashi, and N.~Imoto,
% ``Configuration of Separability and Tests for Multipartite Entanglement in Bell-Type Experiments''
  Phys. Rev. Lett.{\bf 89}, 260401 (2002).
%  doi:10.1103/PhysRevLett.89.260401

%\cite{Yu}
\bibitem{Yu}  
 S.~Yu,  Z.~B.~Chen,   J.~W.~Pan, and Y.~D.~Zhang, 
% ``Classifying $N$-Qubit Entanglement via Bell's Inequalities''
 Phys. Rev. Lett. 90, 080401 (2003).
 %  doi:10.1103/PhysRevLett.90.080401
[arXiv:quant-ph/0211063]

%\cite{Campo:2005sv}
\bibitem{Campo:2005sv} 
  D.~Campo and R.~Parentani,
  %``Inflationary spectra and violations of Bell inequalities,''
  Phys.\ Rev.\ D {\bf 74}, 025001 (2006)
 % doi:10.1103/PhysRevD.74.025001
  [astro-ph/0505376].\\
  %%CITATION = doi:10.1103/PhysRevD.74.025001;%%
  D.~Campo and R.~Parentani,
  %``Quantum correlations in inflationary spectra and violation of bell inequalities,''
  Braz.\ J.\ Phys.\  {\bf 35}, 1074 (2005)
 % doi:10.1590/S0103-97332005000700016
  [astro-ph/0510445].
  %%CITATION = doi:10.1590/S0103-97332005000700016;%%
  
%\cite{Maldacena:2012xp}
\bibitem{Maldacena:2012xp} 
  J.~Maldacena and G.~L.~Pimentel,
  %``Entanglement entropy in de Sitter space,''
  JHEP {\bf 1302}, 038 (2013)
%  doi:10.1007/JHEP02(2013)038
  [arXiv:1210.7244 [hep-th]].
  %%CITATION = doi:10.1007/JHEP02(2013)038;%%

%\cite{Kanno:2014lma}
\bibitem{Kanno:2014lma} 
  S.~Kanno, J.~Murugan, J.~P.~Shock and J.~Soda,
  %``Entanglement entropy of $\alpha$-vacua in de Sitter space,''
  JHEP {\bf 1407}, 072 (2014)
%  doi:10.1007/JHEP07(2014)072
  [arXiv:1404.6815 [hep-th]].
  %%CITATION = doi:10.1007/JHEP07(2014)072;%%

%\cite{Iizuka:2014rua}
\bibitem{Iizuka:2014rua} 
  N.~Iizuka, T.~Noumi and N.~Ogawa,
  %``Entanglement entropy of de Sitter space $\alpha$-vacua,''
  Nucl.\ Phys.\ B {\bf 910}, 23 (2016)
  %doi:10.1016/j.nuclphysb.2016.06.024
  [arXiv:1404.7487 [hep-th]].
  %%CITATION = doi:10.1016/j.nuclphysb.2016.06.024;%%

%\cite{Kanno:2016qcc}
\bibitem{Kanno:2016qcc} 
  S.~Kanno, M.~Sasaki and T.~Tanaka,
  %``Vacuum State of the Dirac Field in de Sitter Space and Entanglement Entropy,''
  JHEP {\bf 1703}, 068 (2017)
  %doi:10.1007/JHEP03(2017)068
  [arXiv:1612.08954 [hep-th]].
  %%CITATION = doi:10.1007/JHEP03(2017)068;%%

%\cite{Kanno:2014bma}
\bibitem{Kanno:2014bma} 
  S.~Kanno, J.~P.~Shock and J.~Soda,
  %``Entanglement negativity in the multiverse,''
  JCAP {\bf 1503}, no. 03, 015 (2015)
  %doi:10.1088/1475-7516/2015/03/015
  [arXiv:1412.2838 [hep-th]].
  %%CITATION = doi:10.1088/1475-7516/2015/03/015;%%

%\cite{Lim:2014uea}
\bibitem{Lim:2014uea} 
  E.~A.~Lim,
  %``Quantum information of cosmological correlations,''
  Phys.\ Rev.\ D {\bf 91}, no. 8, 083522 (2015)
  %doi:10.1103/PhysRevD.91.083522
  [arXiv:1410.5508 [hep-th]].
  %%CITATION = doi:10.1103/PhysRevD.91.083522;%%

%\cite{Martin:2015qta}
\bibitem{Martin:2015qta} 
  J.~Martin and V.~Vennin,
  %``Quantum Discord of Cosmic Inflation: Can we Show that CMB Anisotropies are of Quantum-Mechanical Origin?,''
  Phys.\ Rev.\ D {\bf 93}, no. 2, 023505 (2016)
 % doi:10.1103/PhysRevD.93.023505
  [arXiv:1510.04038 [astro-ph.CO]].
  %%CITATION = doi:10.1103/PhysRevD.93.023505;%%  
  
%\cite{Kanno:2016gas}
\bibitem{Kanno:2016gas} 
  S.~Kanno, J.~P.~Shock and J.~Soda,
  %``Quantum discord in de Sitter space,''
  Phys.\ Rev.\ D {\bf 94}, no. 12, 125014 (2016)
  %doi:10.1103/PhysRevD.94.125014
  [arXiv:1608.02853 [hep-th]].
  %%CITATION = doi:10.1103/PhysRevD.94.125014;%%

%\cite{Maldacena:2015bha}
\bibitem{Maldacena:2015bha} 
  J.~Maldacena,
  %``A model with cosmological Bell inequalities,''
  Fortsch.\ Phys.\  {\bf 64}, 10 (2016)
 % doi:10.1002/prop.201500097
  [arXiv:1508.01082 [hep-th]].
  %%CITATION = doi:10.1002/prop.201500097;%%
            
%\cite{Choudhury:2016cso}
\bibitem{Choudhury:2016cso} 
  S.~Choudhury, S.~Panda and R.~Singh,
  %``Bell violation in the Sky,''
  Eur.\ Phys.\ J.\ C {\bf 77}, no. 2, 60 (2017)
  %doi:10.1140/epjc/s10052-016-4553-3
  [arXiv:1607.00237 [hep-th]].
  %%CITATION = doi:10.1140/epjc/s10052-016-4553-3;%%

%\cite{Ade:2015lrj}
\bibitem{Ade:2015lrj} 
  P.~A.~R.~Ade {\it et al.} [Planck Collaboration],
  %``Planck 2015 results. XX. Constraints on inflation,''
  Astron.\ Astrophys.\  {\bf 594}, A20 (2016)
  %doi:10.1051/0004-6361/201525898
  [arXiv:1502.02114 [astro-ph.CO]].
  %%CITATION = doi:10.1051/0004-6361/201525898;%%

%\cite{Albrecht:2014aga}
\bibitem{Albrecht:2014aga} 
  A.~Albrecht, N.~Bolis and R.~Holman,
  %``Cosmological Consequences of Initial State Entanglement,''
  JHEP {\bf 1411}, 093 (2014)
  %doi:10.1007/JHEP11(2014)093
  [arXiv:1408.6859 [hep-th]].
  %%CITATION = doi:10.1007/JHEP11(2014)093;%%
    
%\cite{Kanno:2015ewa}
\bibitem{Kanno:2015ewa} 
  S.~Kanno,
  %``A note on initial state entanglement in inflationary cosmology,''
  Europhys.\ Lett.\  {\bf 111}, no. 6, 60007 (2015)
  %doi:10.1209/0295-5075/111/60007
  [arXiv:1507.04877 [hep-th]].
  %%CITATION = doi:10.1209/0295-5075/111/60007;%%

%\cite{Kanno:2015lja}
\bibitem{Kanno:2015lja} 
  S.~Kanno,
  %``Cosmological implications of quantum entanglement in the multiverse,''
  Phys.\ Lett.\ B {\bf 751}, 316 (2015)
%  doi:10.1016/j.physletb.2015.10.050
  [arXiv:1506.07808 [hep-th]].
  %%CITATION = doi:10.1016/j.physletb.2015.10.050;%%
  
%\cite{Collins:2016ahj}
\bibitem{Collins:2016ahj} 
  H.~Collins and T.~Vardanyan,
  %``Entangled Scalar and Tensor Fluctuations during Inflation,''
  JCAP {\bf 1611}, no. 11, 059 (2016)
  %doi:10.1088/1475-7516/2016/11/059
  [arXiv:1601.05415 [hep-th]].
  %%CITATION = doi:10.1088/1475-7516/2016/11/059;%%
      
%\cite{Bolis:2016vas}
\bibitem{Bolis:2016vas} 
  N.~Bolis, A.~Albrecht and R.~Holman,
  %``Modifications to Cosmological Power Spectra from Scalar-Tensor Entanglement and their Observational Consequences,''
  JCAP {\bf 1612}, no. 12, 011 (2016)
  %doi:10.1088/1475-7516/2016/12/011
  [arXiv:1605.01008 [hep-th]].
  %%CITATION = doi:10.1088/1475-7516/2016/12/011;%%

%\cite{GHZ}
\bibitem{GHZ}
  D.~M.~Greenberger, M.~A.~Horne and A.~Zeilinger,
  %``Going Beyond Bell's Theorem,"
  [arXiv:0712.0921 [quant-ph]]


\end{thebibliography}
\end{document}